# Experimental Study of the Interfacial Waves in Horizontal Stratified Gas-liquid Two-phase Flows by Using the Developed Image Processing Technique

Hadiyan Yusuf Kuntoro[1,2*], Akhmad Zidni Hudaya[1], Okto Dinaryanto[1], Deendarlianto[1,2+], Indarto[1]

[1]Department of Mechanical & Industrial Engineering,
Faculty of Engineering, Universitas Gadjah Mada,
Jalan Grafika No. 2, Yogyakarta 55281, Indonesia.
* Email: hadiyan.y.kuntoro@mail.ugm.ac.id

[2]Center for Energy Studies,
Universitas Gadjah Mada,
Sekip K-1A Kampus UGM, Yogyakarta 55281, Indonesia
[+] Email: deendarlianto@ugm.ac.id

*Abstract*— Experimental series of stratified gas-liquid two-phase flows had been carried out in a 26 mm i.d. transparent acrylic horizontal pipe. The study was aimed to determine the interfacial wave characteristics of the flow and to develop a high quality database of it. The longitudinal section of the pipe was used as the reference section of image recording. Air and water were used as the test fluids, flowing co-currently inside the pipe. The flow behavior was recorded by using a high-speed video camera around 5 m in axial distance from the inlet pipe to ensure the fully-developed stratified gas-liquid two-phase flow. To correct the refraction due to the acrylic pipe, a correction box was employed in the visualization test section. The group of stratified smooth and wavy two-phase flows were successfully recorded and classified on the basis of the visualization study from 24 couples of test condition of superficial water and air velocities. Digital image processing technique was then used to perform quantitative analysis and the results were used to evaluate the existing data. In the present study, the image processing technique was performed to obtain the liquid holdup distribution. The present results show satisfactory agreement with previous works.

## I. INTRODUCTION

Gas-liquid two-phase flow widely exists in many industries, for instance in the petroleum, chemical and nuclear industries. In-depth knowledge regarding these phenomena is highly necessary in order to get better information about the effects to the pipelines system. The complex behavior of flow patterns in two-phase flow had been studied by numerous researchers [1-6] and various models to figure-out the phenomena had been proposed, meanwhile there were some discrepancies among each other. Therefore, the validation is needed to get a better understanding of it.

There are several flow patterns of two-phase flow in a horizontal pipe, and one of them is stratified two-phase flow pattern [7]. It is recognized by the liquid phase flows along the bottom of the pipe whereas the gas phase flows co-currently above it. The gas and liquid phases are separated by an interface that can be smooth or wavy. The behavior of this interface is strongly affected by the interaction between gas and liquid flows. This is because the difference of the local velocity between each phase can initiate the waves on the interface [8]. Furthermore, it can initiate the transition from a stratified to a slug flow pattern [9].

There are a lot of available developed techniques and methods to study the phenomena in two-phase flow. Some examples are capacitance sensor [10], wire-mesh sensor [11], constant electric current method (CECM) [12], ultrasonic detection technique [13], impedance method [14], X-ray tomography [15], gamma-densitometry [16] and laser focus displacement meter [17].

Here, the researchers need to solve large quantities of visualization data which contain different information. Hence, the use of digital device which is capable to process them is needed. Image processing technique is a powerful technique to investigate the two-phase flow phenomena because provides a better understanding of important flow phenomena which are difficult to observe by the other methods. This technique had been used in the wide range of applications and studies due to their non-intrusive capability to obtain and process a lot of visualization data which contain many complexities. In the field of two-phase flow studies, image processing technique had been used by Gopal and Jepson [18], Mayor et al. [19], Ozbayoglu and Yuksel [20] and do Amaral et al. [21]. Gopal and Jepson [18] developed the digital image analysis techniques for the study of velocity and void profile in slug flow in a horizontal 75 mm i.d. Plexiglass pipe. Mayor et al. [19] carried out the implementation of digital image processing technique for study of gas-liquid slug flow along vertical pipes. For horizontal annuli cases, Ozbayoglu and Yuksel [20] analyzed the gas-liquid two-phase behavior with image processing techniques. Meanwhile, do Amaral et al. [21] investigated using image processing techniques in two-phase slug flows flowing inside horizontal 26 mm i.d. acrylic pipe. On the other hand, the implementation of digital image processing technique to study the stratified two-phase flow is scarce.

In the present study, the flow visualization and the digital image processing techniques were developed to perform a detail analysis on the local characteristics of stratified flow. In the scientific viewpoint, the lack of a high quality database on the stratified gas-liquid two-phase flow is the motivation of this research.

## II. EXPERIMENTAL SET-UP

### A. Experimental Apparatus

The experimental series were carried out at the Fluid Mechanics Laboratory, Department of Mechanical and Industrial Engineering, Faculty of Engineering, Universitas Gadjah Mada. Water and air were used as liquid and gas fluids moving co-currently along a horizontal pipe.

The stratified gas-liquid two-phase flow pattern was simulated with well-arranged experimental apparatus. In order to ensure the outlet mixing flow of air-water becomes stratified two-phase flow, a flat plate was used to separate the inlet. A transparent acrylic horizontal pipe with internal diameter 26 mm was used in these experimental series to ensure the observation of the gas-liquid interface. The visualization test section pipe length was 1 m. To remove the refraction effect due to the acrylic pipe, a correction box was used (Fig. 1). The stratified gas-liquid two-phase flow characteristics were observed by a high-speed video camera. Canon Power Shot S100 was used as a high-speed video camera with frame rate video recording 120 fps and 640 x 480 pixel resolutions. LED lamp was employed as the light source. The flow was recorded on video for 30 s each matrix data and the images were digitized on a personal computer. The photograph and the schematic diagram of experimental apparatus are shown in Figs. 2 and 3.

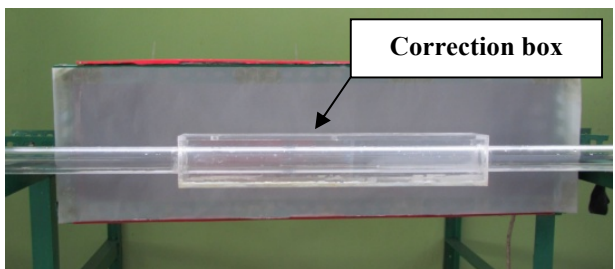

Fig. 1. Visualization test section.

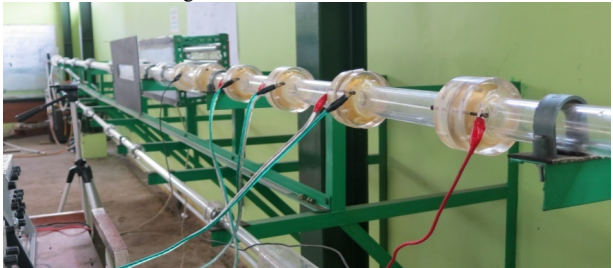

Fig. 2. Experimental apparatus.

### B. Experimental Procedures

The present work is within the research framework to investigate in detail the whole flow patterns behavior in a horizontal pipe and the present paper clarifies on the focus of the stratified flow pattern. Present research matrix data of stratified two-phase flow refer to Mandhane et al. [22]'s flow pattern map of stratified smooth and wavy regions (Fig. 4).

First of all, the inlet tank was filled with water for the first start until 75% of the inlet tank capacity. Before filling the inlet tank, the water was filtered using porous media to avoid dirt. To prevent the water enters the air flowmeter, the water valve was closed and the by-pass valve of the main pump was opened. Next, the compressor was turned on and the air valve was opened to generate pressure in order to prevent the water from entering the air flowmeter. Subsequently, the main pump and the circulation pump were turned on. The air valve was gradually set until reaching a certain discharge. After that, the superficial air velocity ($J_G$) was calculated and noted. By keeping the superficial air velocity ($J_G$) constant, the water valve was gradually opened until the stratified flow pattern appears, and then, the water discharge was measured to calculate the superficial water velocity ($J_L$). The superficial water velocity ($J_L$) was varied from 0.016 m/s until 0.092 m/s. The superficial air velocity ($J_G$) was varied from 1.02 m/s until 3.77 m/s. The variation of $J_G$ and $J_L$ can be seen in Table 1. In the specific $J_G$ and $J_L$, the flow was recorded by using a high-speed video camera 120 fps with 640 x 480 pixel resolutions for 30 s. The recorded video was analyzed quantitatively by using the digital image processing technique on a personal computer.

TABLE I. RESEARCH MATRIX DATA.

| | | $J_L$ (m/s) | | | | | |
|---|---|---|---|---|---|---|---|
| | | 0.016 | 0.031 | 0.047 | 0.063 | 0.077 | 0.092 |
| $J_G$ (m/s) | 1.02 | 1 | 2 | 3 | 4 | 5 | 6 |
| | 1.88 | 7 | 8 | 9 | 10 | 11 | 12 |
| | 2.83 | 13 | 14 | 15 | 16 | 17 | 18 |
| | 3.77 | 19 | 20 | 21 | 22 | 23 | 24 |

## III. IMAGE PROCESSING TECHNIQUE

There are sequential processes to obtain the desired data from the recorded images. The image processing concept is to change the RGB images (from the recorded images) by means of sequential processes to be binary images that only contain two type value of data, 0 for black and 1 for white. From the binary images, the distribution of the liquid holdup data are obtained. Liquid holdup ($\eta$) is defined as the cross sectional area of the liquid phase divided by the cross sectional area of the inner pipe. Here, the plane curvature assumption is used.

The recorded images were processed using MATLAB through algorithm programs with following sequences: High-speed video camera recorded the stratified gas-liquid two-phase flow in the video file format 120 fps and 640 x 480 pixel resolutions for 30 s each matrix data. Then, the recorded video was extracted by the extraction software (VirtualDub) to obtain the sequential individual RGB image (Fig. 5).

Next, each image was converted to grayscale image format (1 matrix layer) by obtaining the red layer in the RGB image and using it as the new matrix layer (the R layer) rather than using the *rgb2gray* function in MATLAB. Ozbayoglu and Yuksel [20] performed this method also, but there was no detail explanation about the reason using this method. Fig. 6 shows the comparison between grayscale images from 4 different conversion methods.

Fig. 6.b shows the most contrast and less noise image rather than the others. Fig. 6.b is suitable for further image processing because the objective is to obtain the gas-liquid interface (Figs. 7.a – 7.b) and eliminate the noise (the undesired recorded image – Figs. 7.c – 7.d). The comparison in Fig. 6 shows that the R layer separation method gives better result of image rather than the *rgb2gray* MATLAB function method.

The noise reduction process includes image complement process (Fig. 7.d.) and image filtering process (Figs. 7.e – 7.f). In the image complement process, the values of image data were reversed to the complement values.

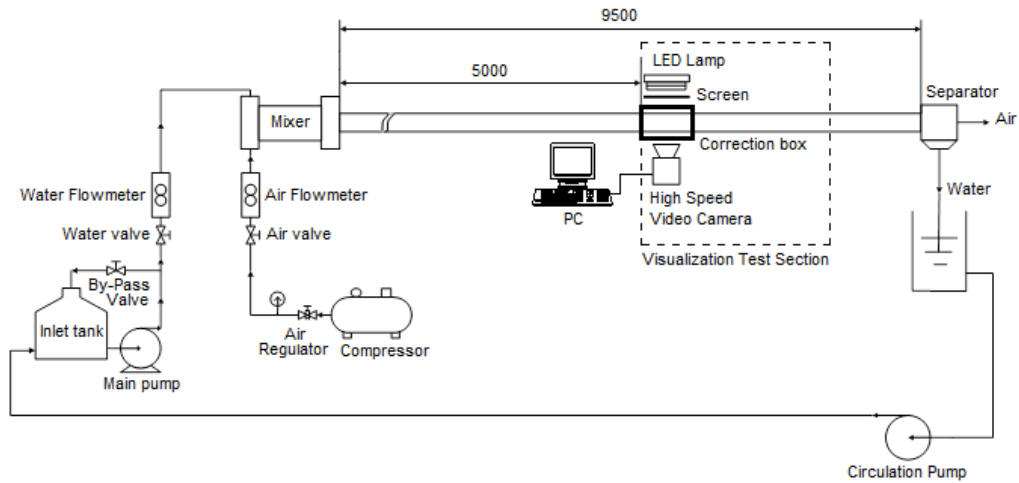

Fig. 3. Schematic diagram of experimental apparatus.

For the image filtering process, zero filtering (Fig. 7.e) and average filtering (Fig. 7.f) were used. The idea of average filtering is to replace each pixel value in an image with the average value of its neighbors, including itself. The purpose is to eliminate the pixel values which are unrepresentative of their surroundings. The 3 x 3 neighbor pixel size was used in this filtering. The present work also developed the new filtering method, called zero filtering. The zero filtering concept is to replace the non-interfacial gas-liquid image to be 0 value (black color). The plot technique should be performed first to separate the matrix image area between the gas-liquid interface and the noise.

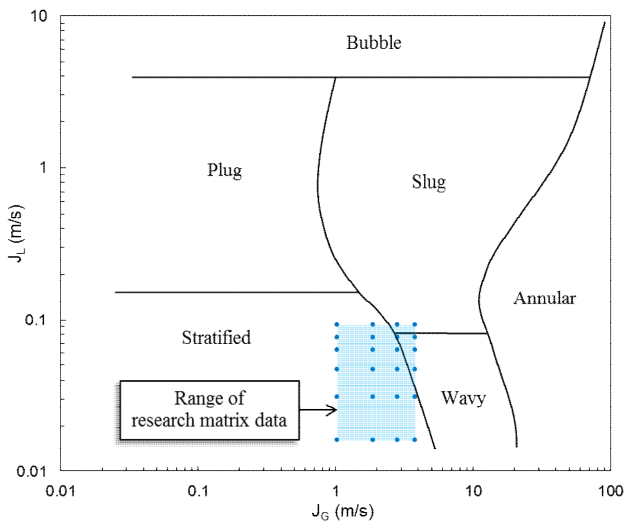

Fig. 4. Range of research matrix data in Mandhane et al. [22]'s flow pattern map.

The image segmentation process includes image contrast enhancement process (Fig. 7.g), thresholding, binary image conversion (Fig. 7.h) and morphological image process (Fig. 7.i). Image contrast enhancement process was used to contrast the gas-liquid interface image against the background image. Thresholding and binary image conversion were done to convert the grayscale images into binary images. For morphological image process, skeletonize method was used. After all sequence processes were done, the skeletonize images were obtained.

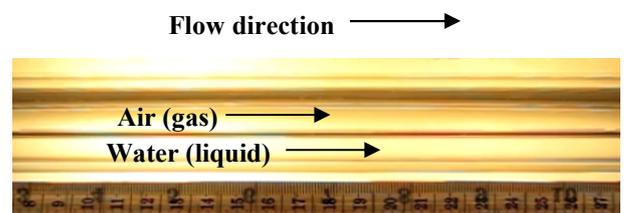

Fig. 5.a.  Research matrix data number 1.

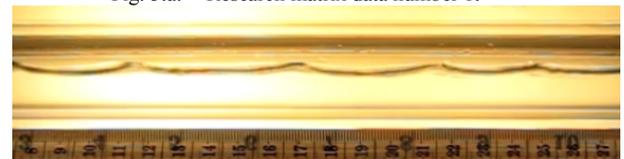

Fig. 5.b.  Reserch matrix data number 6.

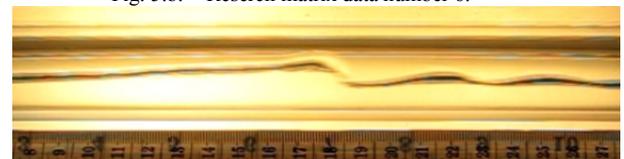

Fig. 5.c.  Research matrix data number 12.

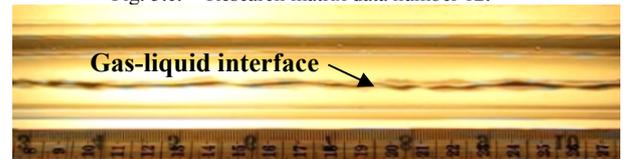

Fig. 5.d.  Research matrix data number 15.

Fig. 5. RGB image samples of research matrix data number: a. 1; b. 6; c. 12; d. 15.

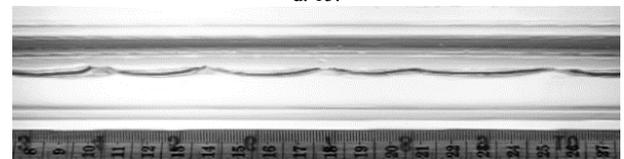

Fig. 6.a.  *rgb2gray* MATLAB function method.

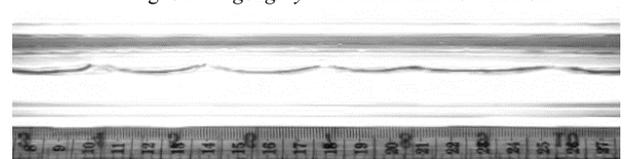

Fig. 6.b.  R layer separation method.

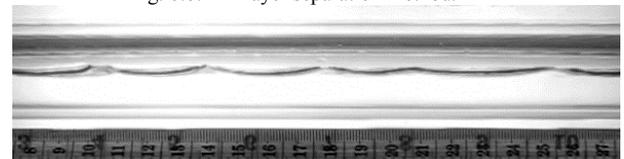

Fig. 6.c.  G layer separation method.

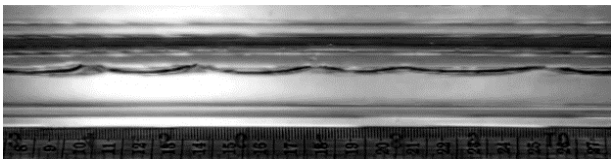

Fig. 6.d.    B layer separation method.
Fig. 6. The comparison between grayscale images from 4 different conversion method.

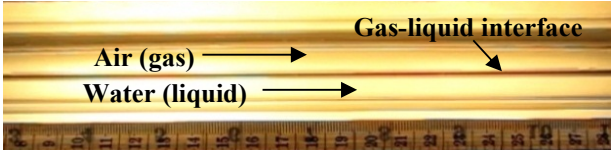

Fig. 7.a.    RGB image.

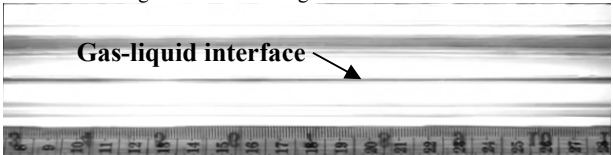

Fig. 7.b.    Grayscale image.

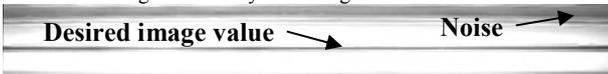

Fig. 7.c.    Image cropping result.

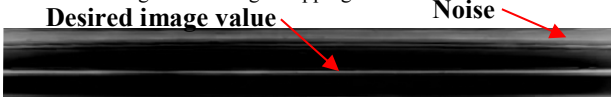

Fig. 7.d.    Image complement result.

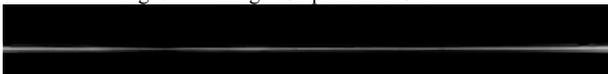

Fig. 7.e.    Image after zero filtering.

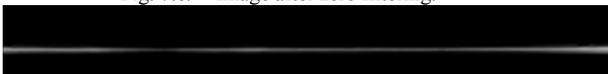

Fig. 7.f.    Image after average filtering.

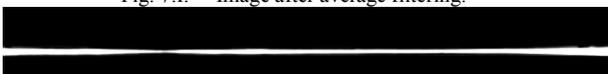

Fig. 7.g.    Image after contrast enhancement process.

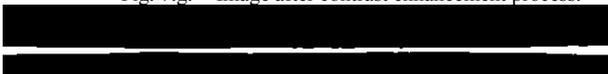

Fig. 7.h.    Binary image.

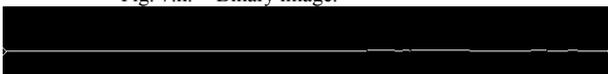

Fig. 7.i.    Image after skeletonize process.
Fig. 7. The results of each sequence in the image processing technique.

## IV.    RESULTS AND DISCUSSIONS

### A.    Flow Visualization

#### 1)    Stratified Smooth

Stratified smooth flow pattern is characterized by the smooth flat undisturbed interface separating the gas and liquid phases due to the significant difference in density. This pattern occurs when the superficial gas and liquid velocities are low enough and not sufficient to cause waves. The inside pipe area is confined by the red box, as shown in Fig. 8.

In the time series graph (Fig. 9.a), the stratified smooth flow pattern can be recognized by the flat plot which indicates the absence of fluctuation of the liquid holdup against the change of time. The gathered dominant distribution of probability distribution function (PDF) plot in specific value is also a characteristic of this flow pattern (Fig. 9.b). This plot denotes that there are not many variations of liquid holdup in the range of a certain time. The absence of fluctuation in the liquid holdup indicates the absence of pressure fluctuation that occurs in the stratified smooth flow pattern.

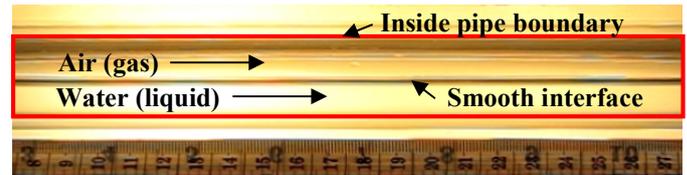

Fig. 8. Stratified smooth ($J_L$ = 0.031 m/s and $J_G$ = 1.02 m/s).

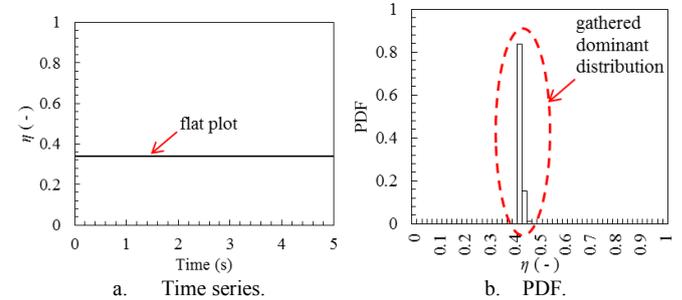

a.    Time series.                              b.    PDF.
Fig. 9. Liquid holdup distribution of stratified smooth ($J_L$ = 0.031 m/s and $J_G$ = 1.02 m/s).

#### 2)    Stratified Ripple

The occurrence of stratified ripple is identified by the presence of the periodic waves in the gas-liquid interface. Previous investigations [7, 22, 23] distinguished the stratified flow pattern only into stratified smooth and stratified wavy. Further investigation by Spedding and Nguyen [24] using more sophisticated method differentiated the stratified wavy into stratified ripple and stratified roll, which were also confirmed in the present work. From the present experimental observations, there were two distinguishable waves occur in the stratified wavy region: stratified ripple and stratified roll.

In the stratified ripple flow pattern, the wave is like "peak and valley" wave flowing periodically. In this wavy flow pattern, the waves occur but the peaks do not reach the upper part of inside pipe, as seen in Fig. 10. These waves are initiated by the gas flow under conditions where the velocity of the gas is sufficient to cause waves to form but slower than that needed for the rapid wave growth which causes transition to intermittent or annular flow [23].

In the time series graph (Fig. 11.a), the stratified ripple shows the fluctuating plot that indicates the fluctuation of liquid holdup against the change of time. The fluctuation of the liquid holdup also indicates the fluctuation of the pressure which occurs inside the pipe.

The distribution value of liquid holdup in the stratified ripple's PDF (Fig. 11.b) looks to spread over a certain range of liquid holdup comparing to the stratified smooth's PDF (Fig. 9.b). This informs that the widened dominant distribution of liquid holdup occurs in the stratified ripple.

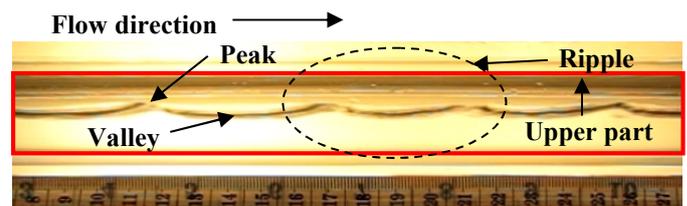

Fig. 10. Stratified ripple ($J_L$ = 0.092 m/s and $J_G$ = 1.02 m/s).

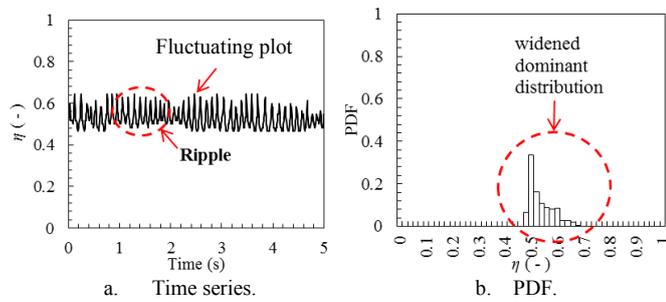

a. Time series.　　　　　　b. PDF.
Fig. 11. Liquid holdup distribution of stratified ripple ($J_L$ = 0.092 m/s and $J_G$ = 1.02 m/s).

*3) Stratified Roll*

Essentially, stratified roll flow pattern is classified into stratified wavy. The stratified wavy itself has two distinguishable wave patterns: the ripple, as discussed in the previous section, and the roll. The roll wave occurs when the shear stress along the flow direction increases and pushes the liquid flow to recede but the liquid discharge is constant, which results in the sudden increase of the liquid holdup and appears as the roll wave (Fig. 12).

The wave structure of the stratified roll is like "whale-head shape" flowing along direction of the flow. It is important to note that the front of the wave is steeper than the back. This interfacial behavior is depicted in the time series graph of the liquid holdup as seen in Fig. 13.a. The sudden increase of the liquid holdup occurs in this flow pattern, so does the pressure. The sudden increase of the liquid holdup also indicates the onset of the initiation of the roll wave.

The stratified roll's PDF (Fig. 13.b) looks similar to the stratified ripple's PDF (Fig. 11.b) which has a widened dominant distribution value of the liquid holdup. Hence, this distribution type is a characteristic of PDF for all stratified wavy two-phase flow patterns.

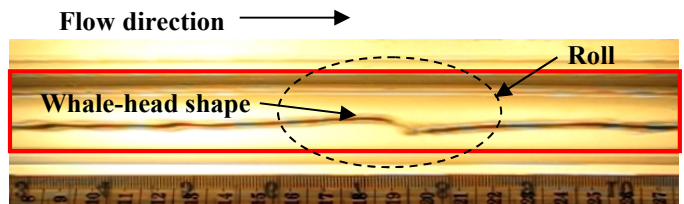

Fig. 12. Stratified roll ($J_L$ = 0.063 m/s and $J_G$ = 2.83 m/s).

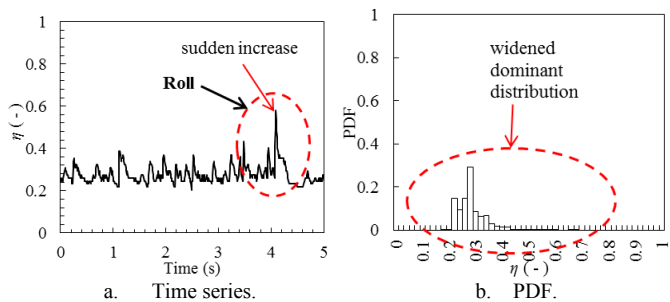

a. Time series.　　　　　　b. PDF.
Fig. 13. Liquid holdup distribution of stratified roll ($J_L$ = 0.063 m/s and $J_G$ = 2.83 m/s).

*4) Pseudo-slug*

A good investigation of the pseudo-slug was performed by Lin and Hanratty [25]. They recorded the flow visualization of it, and the present experiments confirmed their result. Lin and Hanratty [25] also proposed a flow pattern map of the region of the pseudo-slug.

Pseudo-slug is identified by the presence of the hydraulic jump (Fig. 14) which is defined as the sudden wave that touches the upper part of the inside pipe momentarily without causing liquid blockage inside the pipe. The absence of liquid blockage is indicated by no plot in the time series graph (Fig. 15.a) that reaches the value of one which represents fully-liquid in cross section area of the pipe. The collision between the waves and the upper part of the inside pipe enables the atomization/droplets to arise.

Pseudo-slug occurs when the shear stress increases and pushes the liquid flow to recede but the liquid discharge is constant and higher than in the stratified roll. The sudden increase of the liquid holdup in pseudo slug is much steeper than in the stratified roll, as compared Figs. 15.a to 13.a. Similar to the PDF characteristics of the stratified wavy (see Figs. 11.b and 13.b), pseudo-slug's PDF also shows the same distribution type, as seen in Fig. 15.b. Therefore, basing on this reason, some researchers [7, 22, 23] grouped pseudo-slug into stratified wavy. However, the others [25] categorized pseudo-slug into transition flow pattern from stratified to slug or annular flow pattern.

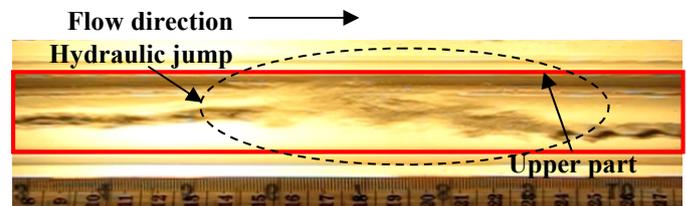

Fig. 14. Pseudo-slug ($J_L$ = 0.077 m/s and $J_G$ = 3.77 m/s).

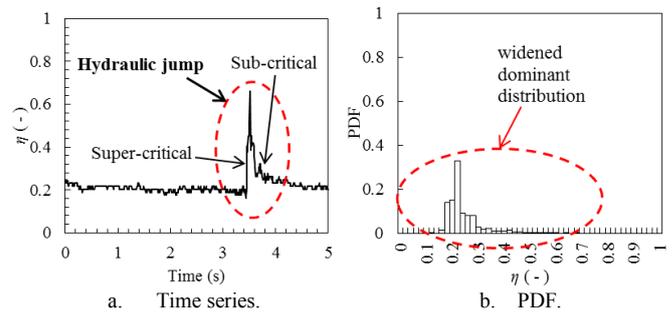

a. Time series.　　　　　　b. PDF.
Fig. 15. Liquid holdup distribution of pseudo-slug ($J_L$ = 0.077 m/s and $J_G$ = 3.77 m/s).

*B. Wave Velocity*

Fig. 16 shows the effect of $J_G$ under a constant $J_L$ on the wave velocity which is calculated from the cross correlation for two successive image points. The wave velocity increases with the $J_G$ which has a significant effect at higher $J_L$.

Fig. 17 shows the effect of $J_L$ under a constant $J_G$ on the wave velocity. As the $J_L$ increases, the wave velocity tends to increase. The increasing of $J_L$ has a significant effect on the increasing of the wave velocity at higher $J_G$.

Franca and Lahey [26] developed a correlation to predict the actual velocity of the gas basing on the drift-flux model for vertical flow. They modified the model so that it is suitable for the case of a horizontal flow, as stated in (1). As seen in Fig. 18, the correlation proposed by Franca and Lahey [26] is in good agreement with the present data.

$$U_G = C_0 J_m + V_{Gj} \qquad (1)$$

$U_G$ is the average gas velocity, $C_0$ is the distribution parameter, $J_m$ is the average volumetric flux and $V_{Gj}$ is the drift velocity.

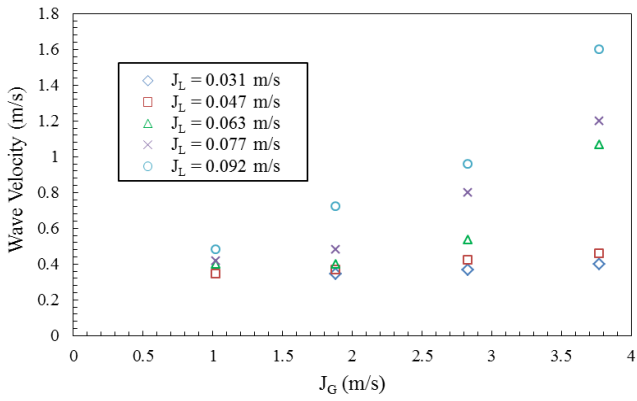

Fig. 16. The effect of $J_G$ on the wave velocity.

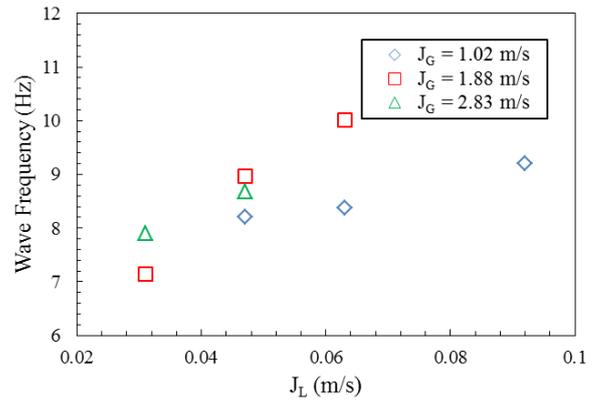

Fig. 19. The effect of $J_L$ on the wave frequency.

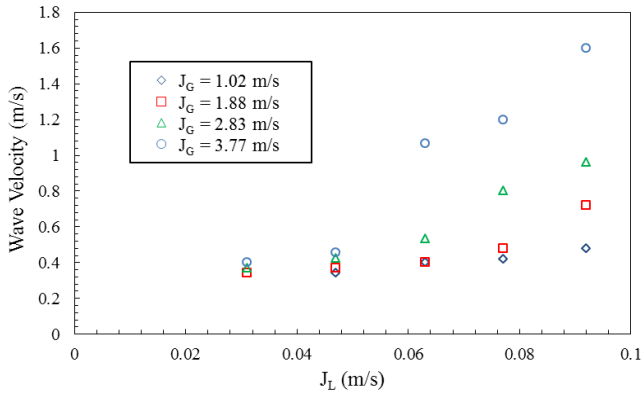

Fig. 17. The effect of $J_L$ on the wave velocity.

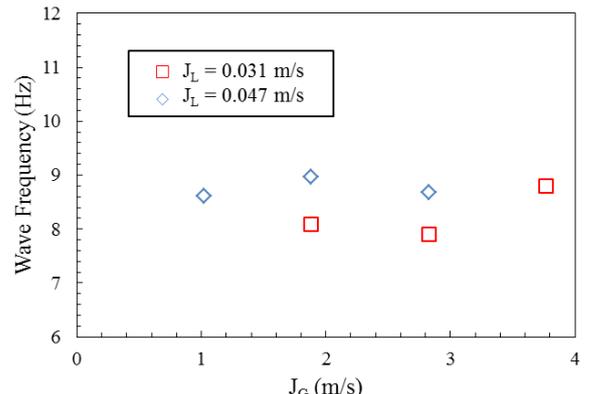

Fig. 20. The effect of $J_G$ on the wave frequency.

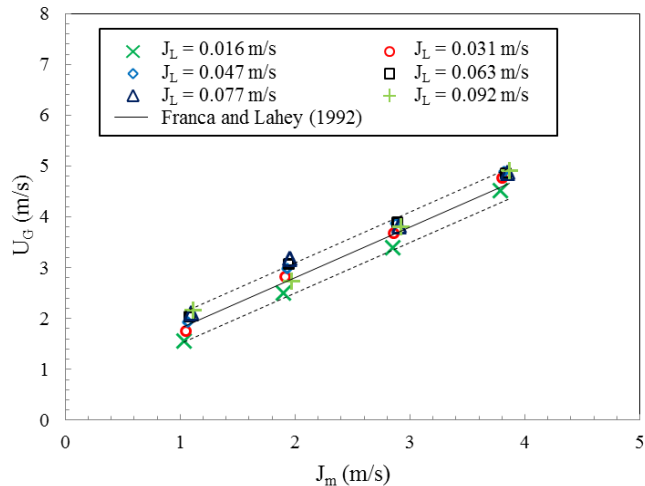

Fig. 18. Gas actual velocity prediction of Franca and Lahey [26].

*C. Wave Frequency*

From the power spectral density analysis, the wave frequency of each stratified flow was studied. The results were plotted against $J_L$ while the $J_G$ was kept constant, and vice versa. As the $J_L$ increases, the wave frequency also increases, as seen in Fig. 19. In comparison with Fig. 19, as shown in Fig. 20, there is no significant change in the wave frequency when the $J_G$ increases at constant $J_L$. This is because the significant difference of air and water densities, so that the little change of $J_L$ has more significant effect comparing to the change of $J_G$ in a same difference value. The wave frequency data in Figs. 19 and 20 are only for flow data that have waves, hence stratified smooth flow pattern is not included in the plot.

V. CONCLUSIONS

Group of stratified gas-liquid two-phase flow pattern in a horizontal pipe has been successfully recorded and is classified into their sub-flow-pattern based on the visual characteristics. The study is performed using digital image processing technique and an algorithm programs of digital image processing in MATLAB to determine the interfacial behavior of stratified gas-liquid two-phase flow is developed with satisfactory result.

There are 4 stratified sub-flow-pattern: stratified smooth, stratified ripple, stratified roll and pseudo-slug. This results confirm the previous observation by Spedding and Nguyen [24]. Basing on the flow visualization, stratified smooth is characterized by the smooth flat undisturbed interface whereas stratified ripple and stratified roll are characterized by the occurrences of waves without reaching the upper part of the inside pipe. The "peak and valley" wave is for stratified ripple and the "whale-head shape" wave is for stratified roll. Pseudo-slug is identified by the presence of the hydraulic jump without causing liquid blockage in the inside pipe. Basing on the time series graph, stratified smooth is recognized by the flat plot and the stratified ripple is recognized by the fluctuating plot. The sudden increase plot occurs in the stratified roll and pseudo-slug. The sudden increase plot is much steeper in the pseudo-slug than in the stratified roll. On the basis of PDF graph, the gathered dominant distribution is shown by the stratified smooth, while the widened dominant distribution is shown by stratified ripple, stratified roll and pseudo-slug which are categorized as the stratified wavy flow pattern.

ACKNOWLEDGMENTS

The authors appreciate the financial support for this project provided by the Indonesia Directorate General for Higher Education (DIKTI), Ministry of Education and Culture, Republic of Indonesia under contract number LPPM-UGM/1448/LIT/2013.

H.Y. Kuntoro acknowledges the Fast Track Scholarship Program from Indonesia Directorate General for Higher Education (DIKTI), Ministry of Education and Culture, Republic of Indonesia.